\documentstyle[12pt]{article}
\begin{document}

\title{{\bf Special Symmetric Quark Mass Matrices}}
\author{\vspace*{1em} \\
{\bf J. I. Silva-Marcos}\thanks{%
E-mail address : juca@nikhef.nl or juca@cfif1.ist.utl.pt} \thanks{On leave from 
the Centro de F\' {\i}sica das Interac\c c\~{o}es Fundamentais (CFIF) 
of the Instituto Superior T\'{e}cnico of Lisbon, Portugal. 
Work supported by the Funda\c c\~{a}o para a Ci\^{e}ncia e Tecnologia  
(FCT) of the Portuguese Governement.}\\
NIKHEF, P.O. Box 41882, 1009 DB Amsterdam, The Nederlands\\
\\
}
\date{}
\maketitle

\begin{abstract}
We give a procedure to construct a special class of symmetric quark mass
matrices near the democratic limit of equal Yukawa couplings for each
sector. It is shown that within appropriate weak-bases, the requirements of
symmetry and arg$[\det (M)]=0$ are very strong conditions,  that necessarily lead to a Cabibbo angle given by $|V_{us}|=%
\sqrt{m_d/m_s}$, and to $|V_{cb}|\sim m_s/m_b$, in first order. In addition,
we prove that the recently classified ans\"atze, which also reproduce these
mixing matrix relations, and which were based on the hypothesis of the
Universal Strength for Yukawa couplings, where all Yukawa couplings have
equal moduli while the flavour dependence is only in their phases, are, in
fact, particular cases of the generalized symmetric quark mass matrix
ans\"atze we construct here. In an excellent numerical example, the experimental values on all quark mixings and masses are accommodated, and the CP
violation phase parameter is shown to be crucially dependent on the values of $m_u$
and $V_{us}$.
\end{abstract}

\setlength{\baselineskip}{14pt} 
\renewcommand{\thesubsection}{\arabic{subsection}} 
\begin{picture}(0,0)
       \put(335,520){NIKHEF 98-008}
       \put(335,505){hep-ph/9807549}
\end{picture}
\vspace{-24pt} \thispagestyle{empty} 


\section{\bf Introduction}

In the Standard Model (SM), the flavour structure of the Yukawa interactions
is not constrained by any symmetry, and the charged currents depend only on
the left handed quark fields. Thus, there is much freedom in defining a
weak-basis for the quarks. This freedom is often used to construct ans\"atze
with which one hopes to find relations between the quark masses and mixings,
and perhaps some day, also find some deeper symmetry beyond the SM, to
restrict the parameters in the Yukawa couplings $\cite{ref0}$.

In the past there have been several attempts at relating the pattern of the
Cabibbo-Kobayashi-Maskawa (CKM) mixing matrix elements to the quark mass
ratios. Many of the ans\"atze, that one finds in the literature, are given
in two distinct types of weak-bases. The well-known Fritzsch ansatz $\cite
{ref1}$, the Hermitic ans\"atze classified by Ramond, Roberts and Ross, and
others $\cite{ref2}$, use the so-called ''heavy'' weak-basis, where one of
the quark mass matrix elements (by consensus usually the (3,3)-element) of
both sectors is much larger than the others. Some of the later may even be zero,
possibly as the result of some (unknown) symmetry.

A different approach is offered by the democratic weak-basis, where the
quark Yukawa interactions become equal and indistinguishable in the limit of
two zero lower masses (for each quark sector apart). Amongst others $\cite
{ref3}$, one of the most interesting examples of this approach is given by
the hypothesis of the Universal Strength for Yukawa couplings (USY) applied
to the quark sector $\cite{ref4}$. In USY all Yukawa couplings have equal
moduli, and the flavour dependence is only in the phases of the Yukawa
coupling matrix elements.

In this paper, we give, in section 2, a procedure to construct a special
class of symmetric ans\"atze near to the democratic quark mass limit. We
obtain this class by demanding that the quark mass matrices be symmetric and
that, for both sectors, arg$[\det (M)]=0$. We then find that (within our
framework) the mixing angles obey the phenomenological mass ratio relations
where $|V_{us}|=(m_d/m_s)^{1/2}$ and $|V_{cb}|$$\sim $$m_s/m_b$. In section 3,
we prove that an important group of recently classified ans\"atze, based on the USY idea, is,
in fact, a particular case within this class. In section 4, a relation is
found between $m_u$, $V_{us}$ and the CP-violation phase. A numerical
example is given, using the most successful of the constructed ans\"atze,
and which accommodates all experimental results on quark masses and mixings.
Finally, in section 5, we present our conclusions.

The purpose of this paper is to find specific ans\"atze by making use of the
freedom that the SM model provides by choosing special weak-bases for the
quarks. As a typical example, we mention the Nearest Neighbour Interaction
(NNI) weak-basis $\cite{ref5}$. It was proven that, without any constraint
on the quark masses and mixings, the mass matrices could be written in the
following way: 
\begin{equation}
\label{eq01}
\begin{array}{ccc}
M_u=\left[ 
\begin{array}{ccc}
0 & a_u & 0 \\ 
a_u^{\prime } & 0 & b_u \\ 
0 & b_u^{\prime } & c_u
\end{array}
\right]  & \qquad ,\qquad  & M_d=\left[ 
\begin{array}{ccc}
0 & a_d & 0 \\ 
a_d^{\prime } & 0 & b_d \\ 
0 & b_d^{\prime } & c_d
\end{array}
\right] 
\end{array}
\end{equation}
The NNI weak-basis allows for different types of ans\"atze $\cite{ref6}$, $%
\cite{ref7}$. By imposing the Hermiticity conditions, i.e., $a_{u,d}^{\prime
}=a_{u,d}^{\star }$, $b_{u,d}^{\prime }=b_{u,d}^{\star }$, $c_{u,d}=$ real,
one obtains the famous Fritzsch ansatz $\cite{ref1}$, now outruled because
of the heavy top mass. Another ansatz, within the NNI weak-basis, can be
constructed if one imposes the non-Hermiticity conditions $a_{u,d}^{\prime
}=a_{u,d}$, $b_{u,d}^{\prime }=c_{u,d}$, which then allow for a heavy top
mass $\cite{ref6}$.

\section{\bf Symmetric ans\"atze}

In this section, we construct a class of complex quark mass matrix
ans\"atze which lead to important mixing matrix relations, where 
$|V_{us}|=(m_d/m_s)^{1/2}$ and $|V_{cb}|\sim m_s/m_b$. The ans\"atze are
symmetric and near to the democratic mass matrix limit. The procedure, we
propose here, is analogous to the NNI-Fritzsh example. This means that we
work in a specific weak-basis, but instead of requiring, e.g., the somewhat
arbitrary Hermiticity conditions, we require that the quark mass matrices be,

- symmetric, and that,

- for each quark sector, arg$[\det (M)]=0$.

The importance of these two requirements, (which will also contribute to the
calculability of the model), for the quark mass matrices is evident:
symmetric fermion mass matrices are crucial in realistic unification
schemes, such as in $SO(10)$ $\cite{ref7a}$, and the requirement that arg$%
[\det (M)]=0$ is a conditio sine qua non for most solutions of the strong CP
problem $\cite{ref8}$.

We begin with the most general, arbitrary, and complex quark mass matrix, 
\begin{equation}
\label{eq18}M_{\circ }=\left[ 
\begin{array}{ccc}
\alpha _{\circ } & \beta _{\circ } & \gamma _{\circ } \\ 
\delta _{\circ } & \epsilon _{\circ } & \zeta _{\circ } \\ 
\eta _{\circ } & \theta _{\circ } & \iota _{\circ }
\end{array}
\right] 
\end{equation}
One can prove that there exists a weak-basis transformation of the
right-handed quark fields, $W$, such that $M_{\circ }\rightarrow M=M_{\circ
}\cdot W$, and where some of the mass matrix elements become equal: 
\begin{equation}
\label{eq19}M=\left[ 
\begin{array}{ccc}
u & u & \widehat{z} \\ \widehat{u} & v & w \\ 
z & \widehat{w} & \widehat{w}
\end{array}
\right] 
\end{equation}
The proof is simple. Take the first line of the general complex quark mass
matrix $M_{\circ }$ in Eq.(\ref{eq18}). This is the (line)vector ${\bf a}%
_{\circ }^{+}=(\alpha _{\circ },\beta _{\circ },\gamma _{\circ })$. Now
construct three orthonormal (column)vectors $({\bf v}_1,{\bf v}_2,{\bf v}_3)$%
, such that the ending points of ${\bf v}_1\,$and ${\bf v}_2$ define a line
which is perpendicular to ${\bf a}_{\circ }^{+}$, then ${\bf a}_{\circ
}^{+}\cdot ({\bf v}_1-{\bf v}_2)=0$ or ${\bf a}_{\circ }^{+}\cdot {\bf v}_1=%
{\bf a}_{\circ }^{+}\cdot {\bf v}_2$. By turning this line around ${\bf a}%
_{\circ }^{+}$, we can assure that the third vector ${\bf v}_3$ lies in such
a direction that another line, through the ending points of ${\bf v}_3$ and $%
{\bf v}_2$, is perpendicular to the third (line)vector of $M_{\circ }$, $%
{\bf c}_{\circ }^{+}=(\eta _{\circ },\theta _{\circ },\iota _{\circ })$.
Thus ${\bf c}_{\circ }^{+}\cdot ({\bf v}_2-{\bf v}_3)=0$ or ${\bf c}_{\circ
}^{+}\cdot {\bf v}_2={\bf c}_{\circ }^{+}\cdot {\bf v}_3$. Therefore,
defining $W=[{\bf v}_1,{\bf v}_2,{\bf v}_3]$, we get the result of Eq.(\ref
{eq19}). Furthermore, one can choose the phase of the complex number $u$ in
such a way that $u(z-w)^2=$real. Up to this point, no conditions were
imposed on $M$ which constrain the mass spectrum or mixing angles. We stress
that the form of $M$ in Eq.(\ref{eq19}) is just a choice of weak-basis.

Now, we demand the first of the two conditions: symmetry. Then $\widehat{u}=u
$, $\widehat{z}=z$, $\widehat{w}=w$, and $M$ becomes of the form, 
\begin{equation}
\label{eq21}M=\left[ 
\begin{array}{ccc}
u & u & z \\ 
u & v & w \\ 
z & w & w
\end{array}
\right] 
\end{equation}
Next, we implement the second condition, arg$[\det (M)]=0$. Computing the
determinant of $M$, 
\begin{equation}
\label{eq22}\det (M)=-u(z-w)^2+(v-u)(uw-z^2)
\end{equation}
and using $u(z-w)^2=$real, we find that $v=u$ is a solution, because $\tan ($%
arg$[\det (M)])$ is proportional to $|v-u|$. Therefore, requiring symmetry
and arg$[\det (M)]=0$ reduces a general complex mass matrix in the
weak-basis of Eq.(\ref{eq19}) to a matrix of the form, 
\begin{equation}
\label{eq23}M=\left[ 
\begin{array}{ccc}
u & u & z \\ 
u & u & w \\ 
z & w & w
\end{array}
\right] =u\quad \left[ 
\begin{array}{ccc}
1 & 1 & z^{\prime } \\ 
1 & 1 & w^{\prime } \\ 
z^{\prime } & w^{\prime } & w^{\prime }
\end{array}
\right] 
\end{equation}
where $z^{\prime }=z/u$ and $w^{\prime }=w/u$.

In the following, we prove that the correct quark mass hierarchy implies that the ansatz of Eq.(\ref{eq23}) is near to the
democratic limit, thus $z^{\prime },w^{\prime }\approx 1$, and that it reproduces the crucial mixing matrix relations, where $%
|V_{us}|=(m_d/m_s)^{1/2} $ and $|V_{cb}|\sim m_s/m_b$. In order to do this,
we introduce the dimensionless square mass matrix $H=3(~M\cdot M^{\dagger
})/tr(~M\cdot M^{\dagger })$. It is clear that by construction $tr(H)=3$.
Taking the right-handed side of Eq.(\ref{eq23}), (we drop the primes on $%
z^{\prime }$ and $w^{\prime }$ to simplify the notation), we get, 
\begin{equation}
\label{eq24}H=\frac 1{~t_{zw}}\cdot \left[ 
\begin{array}{ccc}
2+|z|^2 & 2+zw^{\star } & z^{\star }+w^{\star }+zw^{\star } \\ 
2+z^{\star }w & 2+|w|^2 & z^{\star }+w^{\star }+|w|^2 \\ 
z+w+z^{\star }w & z+w+|w|^2 & |z|^2+2|w|^2 
\end{array}
\right] 
\end{equation}
where $t_{zw}=(4+2|z|^2+3|w|^2)/3$. The determinant $\delta $, the second
invariant $\chi $, and the trace $t$, which are the three invariants of $H$, 
can be expressed in its
dimensionless eigenvalues $\lambda _i$, 
\begin{equation}
\label{eq3b}
\begin{array}{lll}
\delta =\det (H)=\lambda _1\lambda _2\lambda _3 & \qquad \qquad &  \\ 
\chi =\chi (H)=\lambda _1\lambda _2+\lambda _1\lambda _3+\lambda _2\lambda
_3 & \qquad ;\qquad & \lambda _i=3 
\frac{\left( m_i/m_3\right) ^2}{\left[ 1+\left( m_2/m_3\right) ^2+\left(
m_1/m_3\right) ^2\right] } \\ t=tr(H)=\lambda _1+\lambda _2+\lambda _3=3 & 
\qquad \qquad &  
\end{array}
\end{equation}
In order to find the predictions of the ansatz of Eq.(\ref{eq23}), it is
very useful to introduce the following parameterization for $z$ and $w$, 
\begin{equation}
\label{eq25}
\begin{array}{ccc}
z=1+\rho \ e^{i\alpha } & \qquad ,\qquad & w=1+\rho \ e^{i\alpha }-\rho
_{\circ }\ e^{i\beta } 
\end{array}
\end{equation}
This parameterization is general, but we shall now see that 
the mass hierarchy imposes constraints
on $\rho _{\circ }$ and $\rho $, such that these are small. Therefore $%
z,w\approx 1$ and $M$ is almost democratic.

Because of the parametrization of Eq.(\ref{eq25}) one can give a simple
expression for $\rho _{\circ }$ in terms of the determinant $\delta $ of $H$
and $t_{zw}$, 
\begin{equation}
\label{eq26}\rho _{\circ }^2=3\sqrt{3\delta }\cdot \left( \frac{t_{zw}}%
3\right) ^{3/2} 
\end{equation}
where $t_{zw}$ can also be expressed as a function of the variables defined
in Eq.(\ref{eq25}),

\begin{equation}
\label{eq27}t_{zw}=3+\frac{10}3\rho \cos (\alpha )-2\rho _{\circ }\cos
(\beta )+\frac 53\rho ^2-2\rho \rho _{\circ }\cos (\alpha -\beta )+\rho
_{\circ }^2
\end{equation}
Using $\delta $ in Eq.(\ref{eq3b}) in terms of mass ratios, we find that the
parameter $\rho _{\circ }$ is proportional to $(m_1m_2)^{1/2}/m_3$. From $H$
in Eq.(\ref{eq24}), one computes also the second invariant as a function of $%
\rho $, $\rho _{\circ }$, $\cos (\alpha )$ and $\cos (\beta )$. From this,
one deduces an expression for the $\rho $ written in terms of quark mass
ratios, a rest-term, and $t_{zw}$, 
\begin{equation}
\label{eq28}\rho ^2=\frac 94\cdot \chi \cdot \left( \frac{t_{zw}}3\right)
^2\cdot \left[ 1+o_{zw}\right] ^{-1}
\end{equation}
where the rest-term $o_{zw}$ is dependent on powers of $\rho $, $\rho
_{\circ }$, $\rho _{\circ }/\rho $ , $\cos (\alpha )$ and $\cos (\beta )$.
Using the expression for $\chi $ in Eq.(\ref{eq3b}) in mass ratios, we find
that $\rho $ is proportional to $m_2/m_3$. We conclude that $\rho _{\circ }$
and $\rho $ are indeed small. Then from Eq.(\ref{eq27}), we find that $%
t_{zw}\approx 3$. Consequently, it is possible to give a leading order
approximation for $\rho _{\circ }$ and $\rho $: 
\begin{equation}
\label{eq28a}
\begin{array}{ccc}
\rho _{\circ }=3\sqrt{3\ }\frac{\sqrt{m_1m_2}}{m_3} & \qquad ,\qquad  & \rho
=\frac 92\ \frac{m_2}{m_3}.
\end{array}
\end{equation}
A more complete expression for $\rho _{\circ }$ and $\rho $ as a power
series in the mass ratios can be derived from the Eqs.(\ref{eq26}, \ref{eq27}%
, \ref{eq28}) using the method of iteration. Starting with the leading order
approximations, one obtains after a couple of iterations, 
\begin{equation}
\label{eq29}
\begin{array}{l}
\rho _{\circ }=3
\frac{\sqrt{3m_1m_2}}{m_3}\cdot \left[ 1+\frac{15}4\left( \frac{m_2}{m_3}%
\right) \cos (\alpha )-\frac{3\sqrt{3}}{2}\frac{\sqrt{m_1m_2}}{m_3}\cos (\beta
)+\cdots \right]  \\  \\ 
\rho =\frac 92\frac{m_2}{m_3}\cdot \left[ 1-\frac{m_1}{m_2}+\frac 12\left( 
\frac{m_2}{m_3}\right) \cos (\alpha )+\sqrt{3}\frac{\sqrt{m_1m_2}}{m_3}\cos
(\beta )+\cdots \right] 
\end{array}
\end{equation}

Resuming: we have extracted $\rho _{\circ }$ and $\rho $ from the two ($%
\delta $ and $\chi $) mass ratio relations of $H$. The two remaining phase
parameters $\alpha $, $\beta $ will be free. However, from Eq.(\ref{eq29}),
we see that the contribution of $\alpha $ and $\beta $ to $\rho _{\circ }$
and $\rho $ are small.

The next step is to compute the unitary matrix which diagonalizes $H$. The
way to do this, is to introduce the power series of $\rho _{\circ }$, $%
\rho $ in the parametrization of $z$, $w$ of Eq.(\ref{eq25}), and in the
matrix $H$ of Eq.(\ref{eq24}). Thus, we obtain the square matrix $H$ as a
power series in the mass ratios, and it is then easy to calculate the
eigenvectors as a series in these ratios, because the eigenvalues of $H$ are
also expressed as functions of ratios of masses. We prefer, however, to
calculate the diagonalization matrix $V$ in an appropriate ''heavy''
weak-basis for $H$. In this weak-basis all matrix elements of $H$ are small,
except $H_{33}\approx 3$, and only the relevant contributions of $H_u$ and $%
H_d$ to $V_{CKM}$ are present. Thus the irrelevant contributions to the Cabibbo Kobayashi Maskawa mixing matrix $V_{CKM}$, where,

\begin{equation}
\label{eq5b}V_{CKM}=V_u^{\dagger }\cdot V_d 
\end{equation}
and which would cancel out in the matrix product, are absent. In this way, $%
V_u$ and $V_d$ are both near to ${1\>\!\!\!{\rm I}}$. The ''heavy-basis'' is
defined in the following way:

\begin{equation}
\label{eq5a}
\begin{array}{ccc}
\begin{array}{l}
H_u\rightarrow H_u^{\rm{Heavy}}=F^{\dagger }\cdot H_u\cdot F \\  
\\ 
H_d\rightarrow H_d^{\rm{Heavy}}=F^{\dagger }\cdot H_d\cdot F 
\end{array}
& \qquad ;\qquad & F=\left[ 
\begin{array}{ccc}
\frac 1{\sqrt{2}} & \frac 1{\sqrt{6}} & \frac 1{
\sqrt{3}} \\ \frac{-1}{\sqrt{2}} & \frac 1{\sqrt{6}} & \frac 1{
\sqrt{3}} \\ 0 & \frac{-2}{\sqrt{6}} & \frac 1{\sqrt{3}} 
\end{array}
\right] 
\end{array}
\end{equation}
We find, in leading and next leading order, 
\begin{equation}
\label{eq30}
\begin{array}{rr}
|V_{12}|=\sqrt{\frac{m_1}{m_2}}\left[ 1-\frac{m_1}{2m_2}+\frac{m_2}{4m_3}%
\cos (\alpha )\right] ; & |V_{23}|= 
\sqrt{2}\frac{m_2}{m_3}\left[ 1-\sqrt{\frac{3m_1}{4m_2}}\cos (\alpha -\beta
)\right] \\  &  \\ 
|V_{13}|=\frac 1{\sqrt{2}}\frac{\sqrt{m_1m_2}}{m_3}\left[ 1+\frac{13m_2}{4m_3%
}\cos (\alpha )\right] ; & |V_{31}|=\frac 3{\sqrt{2}}\frac{\sqrt{m_1m_2}}{m_3%
}\left[ 1-\sqrt{\frac{m_1}{3m_2}}\cos (\alpha -\beta )\right] 
\end{array}
\end{equation}
where we stress again that the phases $\alpha$, $\beta$ are free 
parameters, in the sence that they are not constrained by mass relations. 
This freedom will be used in section 4, where we shall define a specific 
ansatz by fixing these two free phases to accommodate the CP violation 
phase together with the quark masses and mixings.

To complete the description of our special symmetric ans\"atze near the
democratic limit, we contruct another explicit example. As before, we take
the most general complex quark mass matrix of Eq.(\ref{eq18}), but choose a
slightly different weak-basis from the previous one in Eq.(\ref{eq19}). We choose, 
\begin{equation}
\label{eq31}M=\left[ 
\begin{array}{ccc}
u & u & \widehat{z} \\ \widehat{u} & v & v \\ 
z & \widehat{v} & w 
\end{array}
\right] 
\end{equation}
where $u(v-z)^2$ real. We follow the prescribed scheme, i.e., require
symmetry and arg$[\det (M)]=0$, and obtain the ansatz, 
\begin{equation}
\label{eq31a}M=\left[ 
\begin{array}{ccc}
u & u & z \\ 
u & u & u \\ 
z & u & w 
\end{array}
\right] =u\ \ \ \left[ 
\begin{array}{ccc}
1 & 1 & z^{\prime } \\ 
1 & 1 & 1 \\ 
z^{\prime } & 1 & w^{\prime } 
\end{array}
\right] 
\end{equation}
where again $z^{\prime }=z/u$ and $w^{\prime }=w/u$. The procedure to solve
this ansatz is as in the first example. However, here a different but
general parameterization for $z$ and $w$ is better suited, (we mean in fact the $%
z^{\prime }$ and $w^{\prime }$ on the right-handed side of Eq.(\ref{eq31a})
but as before, in the following, leave out the prime). In order to obtain
similar relations as in Eqs.(\ref{eq26}, \ref{eq28}), we propose, 
\begin{equation}
\label{eq32}
\begin{array}{ccc}
z=1+\rho _{\circ }\ e^{i\beta } & \qquad ,\qquad & w=1+\rho _{\circ }\
e^{i\beta }+\rho \ e^{i\alpha } 
\end{array}
\end{equation}
and find that the diagonalization matrix elements $V_{23}$, $V_{13}$ and $%
V_{31}$, (again in the ''heavy weak-basis'' of Eq.(\ref{eq5a})), are
somewhat different from the previous ansatz. We obtain in leading and next leading order,
\begin{equation}
\label{eq31b}
\begin{array}{rr}
|V_{12}|=\sqrt{\frac{m_1}{m_2}}\left[ 1-\frac{m_1}{2m_2}-\frac{m_2}{4m_3}%
\cos (\alpha )\right] ; & |V_{23}|=\frac 1{
\sqrt{2}}\frac{m_2}{m_3}\left[ 1+\sqrt{\frac{3m_1}{m_2}}\cos (\alpha -\beta
)\right] \\  &  \\ 
|V_{13}|=\frac 1{\sqrt{2}}\frac{\sqrt{m_1m_2}}{m_3}\left[ 1+\frac{5m_2}{4m_3}%
\cos (\alpha )\right] ; & |V_{31}|=\sqrt{\frac 32}\frac{m_1}{m_3}\left[ 1-%
\sqrt{\frac{m_1}{3m_2}}\cos (\alpha -\beta )\right] 
\end{array}
\end{equation}

By now it should be clear how to obtain similiar symmetric ans\"atze near
the democratic limit.

\section{\bf The particular case of USY}

Next, we shall apply the procedure of the previous section to an important
case, where the quark mass matrices are based on the hypothesis of
a Universal Strength for Yukawa couplings. In particular, 
in this section we prove that ans\"atze of USY, thus obtained, coincide
with other 
special USY-ans\"atze, which were classified recently, in
Ref.$\cite{ref9}$, using different arguments of calculability, and for
which all 
parameters are given by the quark mass ratios. These also
predicted $|V_{us}|=(m_d/m_s)^{1/2}$and $|V_{cb}|\sim m_s/m_b$.

In USY, it is assumed that there is only one universal Yukawa strength $%
\lambda $ for all quarks. Two different Higgs doublets $\Phi _u$, $\Phi _d$
give mass to the up and down quarks respectively, and all flavour dependence
is in the phases of the Yukawa couplings. The quark mass matrices have the
following form: 
\begin{equation}
\label{eq1}
\begin{array}{ccc}
M_{u\ =\ }c_u\ \ [\ e^{i\phi _{ij}^u}\ ] & \qquad ,\qquad & M_{d\ =\ }c_d\ \
[\ e^{i\phi _{ij}^d}\ ] 
\end{array}
\end{equation}
with $c_u=\lambda \ $v$_u$, $c_d=\lambda \ $v$_d$, where v$_u=\ <\Phi _u>$, v%
$_d=\ <\Phi _d>$\footnote{%
In our original paper \cite{ref4}, we did not discuss how  a USY mass matrix  could be obtained. This work was done recently by Fishbane and Hung with a mimimum of six Higgs fields \cite{ref12a}}.

We apply our scheme to the USY mass matrix. As can be seen from Eq.(\ref{eq1}%
), the general USY matrix has 9 parameter-phases for each sector. However,
one can choose some of the phases in Eq.(\ref{eq1}) to be equal, in the same
way as it was done for the matrix elements of the 
general case of  Eq.(\ref{eq19}). Then with the
symmetry and arg$[\det (M)]=0$ procedure, we 
obtain a symmetric mass matrix of the following form: 
\begin{equation}
\label{eq12}
\begin{array}{c}
M=c_{\circ }\quad \left[ 
\begin{array}{ccc}
e^{ia} & e^{ia} & e^{-i(a+c)} \\ 
e^{ia} & e^{ia} & e^{ic} \\ 
e^{-i(a+c)} & e^{ic} & e^{ic} 
\end{array}
\right] =c_{\circ }^{\prime }~\quad \left[ 
\begin{array}{ccc}
1 & 1 & e^{iq} \\ 
1 & 1 & e^{i(q-r)} \\ 
e^{iq} & e^{i(q-r}) & e^{i(q-r)} 
\end{array}
\right] 
\end{array}
\end{equation}
where $c_{\circ }^{\prime }=c_{\circ }\ e^{ia}$, $q=-2a-c$, $r=-a-2c$.
Comparing Eq.(\ref{eq12}) with Eq.(\ref{eq23}), we see
that this USY ansatz is a
special case, where general complex numbers have 
been replaced by complex numbers of modulus one. With regard to the
parameter space, we have, applying the parameterization of the general case, given in
Eq.(\ref{eq25}) for $z$ and $w$ in terms of $\rho $, $\rho _{\circ }$, $%
\alpha $ and $\beta $, to this USY ansatz, 
\begin{equation}
\label{eq40}
\begin{array}{lllll}
z=1+\rho \ e^{i\alpha }=e^{iq} &  & \rho =2\ |\sin (\frac q2)| & , & \alpha
=\pm \frac \pi 2+\frac q2 \\  
& \Longrightarrow &  &  &  \\ 
w=1+\rho \ e^{i\alpha }-\rho _{\circ }\ e^{i\beta }=e^{i(q-r)} &  & \rho
_{\circ }=2\ |\sin (\frac r2)| & , & \beta =\pm \frac \pi 2-\frac{r-2q}2 
\end{array}
\end{equation}
where the sign for the phases $\alpha $ and $\beta $ depend on the sign of $q$
and $r$ respectively. 

In principle, one can write the mass power series for the
parameters $q$ and $r$, derived from the series of $\rho $ and $\rho _{\circ
}$ as given in Eq.(\ref{eq29}). However, for this USY-ansatz there exist
exact formul\ae\ for $q$ and $r$ in term of $\delta $ and $\chi $, the mass
ratio invariants of Eq.(\ref{eq3b}): 
\begin{equation}
\label{eq4}
\begin{array}{ccc}
\sin ^2(\frac r2)=\frac 34\sqrt{3\delta } & \qquad ,\qquad & \sin ^2(\frac
q2)=\frac{\frac 9{16}\ \chi -\frac 98\sqrt{3\delta }}{1-\frac 34\sqrt{%
3\delta }} 
\end{array}
\end{equation}
These exact relations are only possible because for the USY case, e.g., the function $%
t_{zw}$, related to the trace of $H$ in Eq.(\ref{eq27}), becomes very
simple, and is equal to $3$. One obtains, of course, the same leading order
approximation relations as in Eq.(\ref{eq28a}): 
\begin{equation}
\label{eq4a}
\begin{array}{ccc}
|r|=3\sqrt{3\ }\frac{\sqrt{m_1m_2}}{m_3} & \qquad ,\qquad & |q|=\frac 92\ 
\frac{m_2}{m_3}. 
\end{array}
\end{equation}

Next, we show that the USY ansatz of Eq.(\ref{eq12}) is
equivalent to one of the ans\"atze of Ref.$\cite{ref9}$. In fact, the
expressions of Eq.(\ref{eq4}), for the phases $r$ and $q$, were given in Ref.$%
\cite{ref9}$ with regard to a different USY-ansatz: 
\begin{equation}
\label{eq12a}M=c_{\circ }\quad \left[ 
\begin{array}{ccc}
1 & e^{ir} & 1 \\ 
e^{iq} & 1 & e^{i(q-r)} \\ 
1 & 1 & 1 
\end{array}
\right] 
\end{equation}
which apparently does not correspond to our USY ansatz in Eq.(\ref{eq12}),
obtained with the symmetry and arg$[\det (M)]=0$ argument. What is the
connection? Writing,

\begin{equation}
\label{eq12b}M=c_{\circ }~\quad \left[ 
\begin{array}{ccc}
1 & e^{ir} & 1 \\ 
e^{iq} & 1 & e^{i(q-r)} \\ 
1 & 1 & 1 
\end{array}
\right] =c_{\circ }\quad P_{23}\cdot \left[ 
\begin{array}{ccc}
1 & 1 & e^{iq} \\ 
1 & 1 & e^{i(q-r)} \\ 
e^{iq} & e^{i(q-r}) & e^{i(q-r)} 
\end{array}
\right] \cdot K_R\cdot P_{23} 
\end{equation}
where $K_R=$diag$(1,1,e^{-i(q-r)})$, and $P_{23}$ is the permutation of the
second with the third quark field, it becomes obvious that our USY-ansatz 
in Eq.(\ref{eq12}), and the USY ansatz of Ref.$\cite{ref9}$ in
Eq.(\ref{eq12a}), are, in fact, equivalent. This is because these ans\"atze are
related by a weak-basis transformation as in Eq.(\ref{eq12b}). 

The diagonalization matrix elements for this USY ansatz can be read off from
the matrix elements for the corresponding general case in Eq.(\ref{eq30}),
using the specific USY phases $\alpha =\pm \pi /2+q/2$ and $\beta =\pm \pi
/2-(r-2q)/2$. The diagonalization matrix elements, that one obtains in this
way are, of course, the same as in Ref.$\cite{ref9}$.

To complete our discussion of the particular hypothesis of USY, we give a
second USY example, derived in a similar way as the previous one, 
and which is related to the second general case in Eq.(\ref{eq31a}): 
\begin{equation}
\label{eq16}
\begin{array}{c}
M=c_{\circ }\quad \left[ 
\begin{array}{ccc}
e^{ia} & e^{ia} & e^{-i2a} \\ 
e^{ia} & e^{ia} & e^{ia} \\ 
e^{-i2a} & e^{ia} & e^{ic} 
\end{array}
\right] =c_{\circ }^{\prime }\quad \left[ 
\begin{array}{ccc}
1 & 1 & e^{ip} \\ 
1 & 1 & 1 \\ 
e^{ip} & 1 & e^{iq} 
\end{array}
\right] 
\end{array}
\end{equation}
where $c_{\circ }^{\prime }=c_{\circ }\ e^{ia}$, $p=-3a$, $q=c-a$. Again, in Ref.$%
\cite{ref9}$ this ansatz was given in a different weak-basis. One can show
the equivalence between the two, by explicitly writing the weak-basis
relation: 
\begin{equation}
\label{eq13}
\begin{array}{c}
M=c_{\circ }\quad \left[ 
\begin{array}{ccc}
1 & 1 & e^{ip} \\ 
1 & 1 & 1 \\ 
e^{ip} & 1 & e^{iq} 
\end{array}
\right] =c_{\circ }\quad P_{13}\cdot \left[ 
\begin{array}{ccc}
e^{ip^{\prime }} & e^{-iq^{\prime }} & 1 \\ 
e^{iq^{\prime }} & 1 & 1 \\ 
1 & 1 & 1 
\end{array}
\right] ^{{\rm {Ref.\cite{ref9}}}}\cdot P_{321}\cdot K_R 
\end{array}
\end{equation}
where $p^{\prime }=q-p$, $q^{\prime }=-p$, the $P$'s are self-evident
permutations, and the phase unitary matrix $K_R=$diag$(1,1,e^{ip})$. The
diagonalization matrix elements are deduced in the same way as in the
previous USY example.

Finally, one can prove that for each USY case in Ref.$\cite{ref9}$ there
exists a corresponding general ansatz, which is obtained with our symmetry
and 
arg$[\det (M)]=0$ scheme.

\section{\bf CP-violation and a numerical example}

In this section, we analise the CP-violation for a typical case of the ans\"atze that we
constructed, and give a numerical example. From the Refs.$\cite{ref9}$, $\cite
{ref10}$ we already know that the particular case of USY can accommodate the
quark masses and mixings.

There are, however, two difficulties with the USY cases of Ref.$\cite{ref9}$. The first is related
to $V_{us}$. For these USY cases, one can choose the phases in $V_{CKM}$
in such a way
that, in leading order, $V_{cs}$, $V_{ub}$ and $V_{cb}$ are real, while, 
\begin{equation}
\label{eq41}V_{us}=\sqrt{\frac{m_d}{m_s}}\pm \sqrt{\frac{m_u}{m_c}} 
\end{equation}
where the sign is a matter of choice, 
dependent on the specific USY sign 
of the phase-parameters as explained for Eq.(\ref{eq40}). 
If one combines the
experimental limits on $m_d/m_s$, $m_s$, $m_c$, then the experimental value
for $|V_{us}|=0.2205(18)$ can only be accommodated if one takes a very small
value for $m_u\leq 1\ MeV$ or even $m_u=0$.

The second problem has to do with $J_{CP}=Im(V_{us}V_{cb}V_{cs}^{\star
}V_{ub}^{\star })$, which measures the CP violation $\cite{ref11}$. In the
USY ans\"atze of the Ref.$\cite{ref9}$, with the phase
convention given above, only second and higher order terms of $V_{cs}$, $%
V_{ub}$, $V_{cb}$ and $V_{us}$ are complex and contribute to $J_{CP}$. The
reason for this shall be made explicitly clear, but essentially this is
because, in the USY cases mentioned, all phase parameters, which enter in the mass matrix
elements, are small, thus, a mass matrix element like $e^{iq}$ in the ansatz
of Eq.(\ref{eq12}), is in leading order equal to $1+iq$. The mass matrix $M$
is, therefore, in this order, equal to the democratic mass matrix plus a
small imaginary matrix. The diagonalization matrix elements $V_{12}$, $%
V_{22} $, $V_{23}$ and $V_{13}$, of both quark sectors get the same phase
factor which cancels out in the CKM matrix product, and only higher order
terms give a contribution to $J_{CP}$. One finds $%
|J_{CP}|=o(10^{-6}-10^{-7}) $, and it is impossible to obtain a large value
for $|J_{CP}|$.

In fact, the two problems are related\footnote{%
Friztsch pointed this out in the context of a different anstaz \cite{ref12}}. If it were
possible to exchange the $\pm $ sign in the expression for $V_{us}$ of Eq.(%
\ref{eq41}) for a phase factor $e^{i\delta }$, while at the same time
keeping $V_{cs}$, $V_{ub}$ and $V_{cb}$ real, then one would solve both
problems. The new relation,

\begin{equation}
\label{eq42}|V_{us}|=\left| \sqrt{\frac{m_d}{m_s}}+e^{i\delta }\cdot \sqrt{%
\frac{m_u}{m_c}}\right| 
\end{equation}
would support a larger value for $m_u$, and one would find in leading order
for $|J_{CP}|$, supposing that $|\sin (\delta )|$ is large, 
\begin{equation}
\label{eq43}|J_{CP}|=|V_{us}V_{cb}V_{cs}V_{ub}|\cdot |\sin (\delta )| 
\end{equation}

The solution to both problems lies, therefore, in choosing a different
complex phase content for the mass matrices of the two quark sectors. Let us
specify more precisely how to obtain this. First we have to choose a
specific ansatz. We find it appealing to have similar ans\"atze for the down
as well as for the up quarks, and propose (the already discussed) mass
matrices of the form: 
\begin{equation}
\label{eq43a}
\begin{array}{ccc}
M_u=c_u\quad \left[ 
\begin{array}{ccc}
1 & 1 & z_u \\ 
1 & 1 & w_u \\ 
z_u & w_u & w_u 
\end{array}
\right] & \qquad ,\qquad & M_d=c_d\quad \left[ 
\begin{array}{ccc}
1 & 1 & z_d \\ 
1 & 1 & w_d \\ 
z_d & w_d & w_d 
\end{array}
\right] 
\end{array}
\end{equation}

In order to find out what the contribution of the CP-phase is to $V_{us}$,
we write the diagonalization equation in the following form: $V\cdot D\cdot
V^{\dagger }=H$, (the indices \{u,d\}have been dropped to simplify), where $%
H $ is given in the ''heavy-basis'' of Eq.(\ref{eq5a}), and $D=$diag$%
(\lambda _1,\lambda _2,\lambda _3)$; the $\lambda _i$ are the eigenvalues of 
$H$ as given in Eq.(\ref{eq3b}). One obtains, thus, (also using the unitarity of $V)$%
: 
\begin{equation}
\label{eq43b}
\begin{array}{l}
(\lambda _2-\lambda _1)\cdot V_{12}V_{32}^{\star }+(\lambda _3-\lambda
_1)\cdot V_{13}V_{33}^{\star }=H_{13} \\ 
(\lambda _2-\lambda _1)\cdot V_{22}V_{32}^{\star }+(\lambda _3-\lambda
_1)\cdot V_{23}V_{33}^{\star }=H_{23} \\ 
(\lambda _2-\lambda _1)\cdot V_{12}V_{22}^{\star }+(\lambda _3-\lambda
_1)\cdot V_{13}V_{23}^{\star }=H_{12} 
\end{array}
\end{equation}
With the general parametrization $z=1+\rho \ e^{i\alpha }$ and $w=1+\rho \
e^{i\alpha }-\rho _{\circ }\ e^{i\beta }$, for $H$ in Eq.(\ref{eq24}), one
finds that $H_{12}$ is an exact real number, while, in first order, $H_{13}=(1/\sqrt{6})\ \rho
_{\circ }\ e^{i\beta }$ and $H_{23}=(2%
\sqrt{2}/3)\ \rho \ e^{i\alpha }$. Therefore, in leading order, the
following expressions, relating the phases $\alpha $ and $\beta $ of the
parameterization of $z$ and $w$ with the phases of $V_{13}$ and $V_{23}$,
hold: 
\begin{equation}
\label{eq43c}
\begin{array}{l}
V_{13}=|V_{13}|\ e^{i\beta } \\ 
V_{23}=|V_{23}|\ e^{i\alpha } 
\end{array}
\end{equation}
where we haved used the possibility to choose the phase of some of the
elements of $V$; in this case $V_{33}=$real. Furthermore, combining the
expressions for $V_{12}$, $V_{23}$ and $V_{13}$ in Eq.(\ref{eq30}), one gets
the leading order relation: $|\lambda _2\cdot V_{12}V_{22}^{\star }|=|\lambda
_3\cdot V_{13}V_{23}^{\star }|$. Then, using $H_{12}=$real, we find with Eq.(\ref
{eq43b}), also in this order, 
\begin{equation}
\label{eq43d}
\begin{array}{lll}
\arg (V_{12}V_{22}^{\star })=-\arg (V_{13}V_{23}^{\star }) & \qquad ,\qquad
& \arg (V_{11}V_{21}^{\star })=-\arg (V_{12}V_{22}^{\star }) 
\end{array}
\end{equation}
where the second relation follows from unitarity.

Finally combining the results for the phases of the matrix elements 
in Eq.(\ref{eq43c}) and Eq.(\ref{eq43d}) for both
sectors $V_d$ and $V_u$, and ignoring higher order contributions, we
obtain for the CP violating phase in $V_{us}$: 
\begin{equation}
\label{eq46}\delta =\pi +(\alpha _d-\beta _d)-(\alpha _u-\beta _u) 
\end{equation}
This equation explains exactly why, for the USY cases of Ref.$\cite{ref9}$, the CP violation is so small: for these
USY cases, the $\alpha $'s and the $\beta $'s are all equal to $\pm \pi /2$, in
leading order (see e.g. Eq.(\ref{eq40})).

With regard to our numerical example, we specify the ansatz in Eq.(\ref{eq43a})
further, by explicitly giving all the phases of the mass matrix elements $%
z_{u,d}$, $w_{u,d}$. We keep the USY ansatz of Eq.(\ref{eq12}) for the down
sector, but introduce for the up sector, matrix elements with moduli that
are not anymore equal to $1$. We propose the following ansatz: 
\begin{equation}
\label{eq45}
\begin{array}{lll}
M_u=c_u\ \left[ 
\begin{array}{ccc}
1 & 1 & z_u \\ 
1 & 1 & w_u \\ 
z_u & w_u & w_u 
\end{array}
\right] 
 & \qquad ;\qquad & M_d=c_d\ \left[ 
\begin{array}{ccc}
1 & 1 & e^{iq_d} \\ 
1 & 1 & e^{i(q_d-r_d)} \\ 
e^{iq_d} & e^{i(q_d-r_d)} & e^{i(q_d-r_d)} 
\end{array}
\right]\end{array}
\end{equation}
where for the up sector: 
\begin{equation}
\label{eq44a}
\begin{array}{ccc}
z_u=1+q_u\ e^{i\pi /3} & \qquad ;\qquad & w_u=1+q_u\ e^{i\pi /3}-r_u 
\end{array}
\end{equation}
The difference in the phases of the matrix elements of the up and down
sectors is evident: 
\begin{equation}
\label{eq44b}
\begin{array}{lll}
\alpha _u=\frac \pi 3 &  & \alpha _d=\pm \frac \pi 2+\frac{q_d}2  \\  
& \qquad ;\qquad &  \\ 
  \beta _u=0 &  &\beta _d=\pm \frac \pi 2-\frac{r_d-2q_d}2
\end{array}
\end{equation}
where the $\pm $ sign for the down sector depends, as was explained, on the
sign of $q_d$ and $r_d$. Remember also that $r_d$ and $q_d$ are small.
Taking different signs for $r_d$ and $q_d$, we get with Eq.(\ref{eq46}) in
leading order (mod $2\pi )$, 
\begin{equation}
\label{eq47}\delta =-\frac \pi 3 
\end{equation}
$\ $

From the Eqs.(\ref{eq42},\ref{eq43}) one would expect that a choice for $%
\delta =\pm \pi /2$ is better suited to support a large $m_u$ value and a
large $|J_{CP}|$. However, numerically one must also adjust the other CKM
matrix elements, and $J_{CP}$ depends not only on the CP-phase $\delta $,
but also on the moduli of $V_{us}$,$V_{cb}$,$V_{cs}$and $V_{ub}$. We have
found the best fit for all experimental results with a CP-phase $\delta
=-\pi /3$, the following parameters and corresponding quark masses:

Input: 
\begin{equation}
\label{eq48}
\begin{array}{ll}
q_u=1.891\times 10^{-2} & r_u=1.074\times 10^{-3} \\ 
q_d=-9.264\times 10^{-2} & r_d=2.205\times 10^{-2} 
\end{array}
\end{equation}
which correspond to mass ratios given by the masses, at $1\ GeV$ $\cite{ref13}$: 
\begin{equation}
\label{eq49}
\begin{array}{ccc}
m_u=3.4\ MeV & m_c=1.43\ GeV & m_t=340\ GeV \\ 
m_d=5.3\ MeV & m_s=135\ MeV & m_b=6.3\ GeV 
\end{array}
\end{equation}

Output: 
\begin{equation}
\label{eq50}\left| V_{CKM}\right| =\left[ 
\begin{array}{lll}
0.9754 & 0.2206 & 0.0029 \\ 
0.2204 & 0.9746 & 0.0395 \\ 
0.0106 & 0.0382& 0.9992 
\end{array}
\right] 
\end{equation}
In this numerical example no approximations were made. The ratio $%
|V_{ub}/V_{cb}|=0.074$ is perfectly well within its experimental limit, $%
|V_{ub}/V_{cb}|^{{\rm {Exp}}}=0.08\pm 0.02$. We get for the CP violation $%
|J_{CP}|=2.02\cdot 10^{-5}$. This value corresponds to a CP-phase $|\sin
(\delta )|=0.816$, which is somewhat smaller ($8\%)$ than the first order prediction $\sin (\pi
/3)=0.886$, because of the influence the higher order contributions to $%
\delta $.

\section{\bf Concluding remarks}

We have shown how to construct a class of symmetric ans\"atze near the
democratic limit, which reproduce the important phenomenological mixing
matrix relations where $|V_{us}|=(m_d/m_s)^{1/2}$ and $|V_{cb}|\sim m_s/m_b$%
. We have proven that the recently classified USY-ans\"atze of Ref.$\cite
{ref9}$, which also reproduce these mixing matrix expressions, are
particular examples within this class. In addition, we have also shown, for
an ansatz-example of the constructed class, how the CP-violation phase can
be computed. For this example, all the experimental values of the
quark masses and mixings, including the CP violation phase, can be accommodated with great success.

We find it very surprising that, with this class, the important issues of
symmetry and arg$[\det (M)]=0$, for the quark mass matrices, become suggestively linked
with the expected phenomenological mixing matrix relations in terms of quark
mass ratios.

\end{document}